\documentclass[reprint,aps,twocolumn,floatfix,superscriptaddress]{revtex4-2}
\usepackage[dvips]{graphicx}
\usepackage{subfigure}
\usepackage{parcolumns}
\usepackage{amsmath}
\usepackage{amssymb}
\usepackage{color}
\usepackage{graphicx}
\usepackage{dcolumn}
\usepackage{bm}
\usepackage{marginnote}

\begin{document}

\preprint{APS/123-QED}

\title{Large Deviations in Continuous Time Random Walks}

\author{Adrian Pacheco-Pozo}
\affiliation{Institut f\"ur Physik, Humboldt Universit\"at zu Berlin, Newtonstra\ss{e} 15, D-12489 Berlin, Germany}
\author{Igor M. Sokolov}
 \affiliation{Institut f\"ur Physik, Humboldt Universit\"at zu Berlin, Newtonstra\ss{e} 15, D-12489 Berlin, Germany}
 \affiliation{IRIS Adlershof, Humboldt Universit\"at zu Berlin, Newtonstra\ss{e} 15, 12489 Berlin, Germany}

\date{\today}

\begin{abstract}
We discuss large deviation properties of continuous-time random walks (CTRW) and present a general expression for the large deviation rate in CTRW in terms of the corresponding rates for the distributions of steps' lengths and waiting times. In the case of Gaussian distribution of steps' lengths the general expression reduces to a sequence of two Legendre transformations applied to the cumulant generating function of waiting times. The discussion of several examples (Bernoulli and Gaussian random walks with exponentially distributed waiting times, Gaussian random walks with one-sided L\'evy and Pareto-distributed waiting times) reveals interesting general properties of such large deviations. 
\end{abstract}

\maketitle


\section{Introduction}

Continuous time random walk (CTRW) introduced in Physics by Montroll and Weiss \cite{Montroll} is a generalization of a simple random walk model, in which the steps follow inhomogeneously in time. However, the precursors of this model can be traced back to the times as early as 1903 in Actuarial Mathematics \cite{Lundberg}  (see \cite{Scalas} for a discussion). 

In the standard variant of the CTRW, times of steps follow a renewal process in which the waiting times for subsequent steps are independent and identically distributed (i.i.d.) random variables. The model is then fully defined by specifying the probability density function (PDF) of waiting times and of spacial displacements in single steps (also being i.i.d. random variables) \cite{KlaSok}. In physics, CTRW is used to model systems showing anomalous diffusion, e.g. situations when the mean square displacements (MSD) $\langle x^2(t) \rangle$ does not grow linearly in time, as predicted by Fick's law, especially in the cases of subdiffusion, when $\langle x^2(t) \rangle \propto t^\alpha$ with $\alpha < 1$. Such anomalous subdiffusion often arises due to the lack of the first moments of the distribution of waiting times. The anomaly in MSD is accompanied by a non-Gaussian shape of PDF of displacements during a given time $t$. This line of modelling, following the pioneering work \cite{Scher}, is discussed in the review articles \cite{Haus} and \cite{Metzler}, with some newer developments discussed in chapters of collective monographs \cite{KlaRadSok} and \cite{KlafLimMetz}. Many recent works have reported weaker type of anomaly, in which the time-dependence of the MSD is linear, but the form of the PDF is pronouncedly non-Gaussian, especially at shorter times. The decay of the tails of the PDFs is often exponential, see e.g. \cite{Chaudhuri,Wang1,Silva} as well as \cite{Chechkin,Barkai,Postnikov} and references therein. In Ref. \cite{Barkai}, Barkai and Burov showed that such exponential decay is a universal behavior of far tails in CTRW, presenting the approach based on subordination of random processes and large deviation theory. In Ref. \cite{Wang}, Wang, Barkai, and Burov presented explicit calculations for the case of exponential and Gamma- (Erlang) distributions of waiting times also using large deviation theory.

The large deviation theory (LDT) is an important mathematical tool, which first arose from the analysis of the insurability and ruin problems, and found its application also for a large class of other problems. In physics, the interest to large deviations is partly due to its intimate connection with the thermodynamical formalism, which leads to additional insights in the thermodynamical limit, and to new results \cite{Touchette}.

In application to the problems of diffusion or anomalous diffusion \cite{Barkai,Wang} the value of the large deviation theory lies in the fact that it adequately describes the behaviors of the PDF in its tails, and also shows where these tails essentially begin, i.e. what part of the distribution is indeed reproduced by the typical approaches aiming onto the adequate description of the PDF close to its mode (like the fluid limits of simple or continuous time random walks). While the very far tails, in the asymptotic domain of LDT, may be hardly accessible to experiments, since the corresponding events might be extremely rare, the knowledge of the transition from central domain to the tail one (i.e. essentially the experimentally accessible pre-asymptotical wing of the distribution) might be of primary importance, especially when the properties of the walks are important for the understanding of the outcomes prone to the influence of rare fluctuations, like the chemical reactions are. 

 In the present work we build on this approach and present a general expression for the large deviations' rate function in CTRW, which we use for discussing several examples beyond the ones considered in \cite{Barkai,Wang}, where we consider both the case when the mean WTD is finite and the case when it diverges. For the case of Gaussian step length distribution, as in \cite{Wang}, the analysis is especially simple, and the general expression reduces to a sequence of two Legendre transformations of a cumulant generating function of waiting times. We note that a considerable work was done in mathematics to investigate large deviation properties of random walks (see e.g. \cite{Kamphorst,Denisov} and references therein) and of CTRW (see e.g. the discussion of the existence and general properties of large deviation rates in Refs. \cite{BoroMogulI,*BoroMogulII,Borovkov2019,*Borovkov2021}), but we concentrate here on physical examples close to the ones discussed in \cite{Barkai,Wang}.

The structure of the present work is as follows: In Sec. \ref{sec:prim} we conduct a preliminary discussion of the model and introduce notation used throughout the work. In Sec. \ref{sec:Main} we derive the main result, which we use in the discussion of several examples; first of them already given in this section. In Sec. \ref{sec:Gauss} we consider the case with Gaussian step length distribution, for which the expressions are especially simple. Here different further examples with and without mean waiting times are considered, which are enough for understanding the physics behind the large deviations in CTRW. Sec. \ref{sec:Sum} concludes the work. 
To make the text self-contained, we present in Appendix \ref{ap:properties}
sketches of the proofs of some standard properties of the rate functions used in the main text, and in Appendix \ref{ap:conexion} the demonstration, on a physical level of rigor, of the known relation between the large deviation rates of 
leading and directing processes of CTRW. Appendix \ref{ap:asymptotic} gives an additional discussion of the asymptotic behavior of Legendre transformations in the case of Pareto distributions supporting intuitive arguments used in the main text.
 
\section{Preliminaries and notation \label{sec:prim}}
Continuous time random walk (CTRW) is a random process which ensues due to subordination of a simple random walk (RW) process $x(n)$ to another random process $n(t)$, 
which gives us the number of steps done by the random walk up to time $t$, and is called a \textit{directing process}, or operational time of the CTRW scheme \cite{KlaSok}. The process $x(n)$ is called with this respect a \textit{parent process} of the CTRW (see Refs. \cite{Gorenflo1,Gorenflo2} for a systematic discussion and terminology). The ``inverse'' process $t(n)$ corresponding to the sum of independent, identically distributed waiting times in CTRW is called a \textit{leading process}. 

Let the PDF of the parent process be $p_n(x)$, and the probability to make exactly $n$ steps up to time $t$ be $\chi_n(t)$. Then the PDF of displacements in CTRW is \cite{KlaSok}
\[
 p(x,t) = \sum_{n=0}^\infty p_n(x) \chi_n(t).
\]
For long $t$, the typical number of steps is very large, and the operational time $\tau=n$ can be taken to be continuous. In this limit
\begin{equation}
 p(x,t) = \int_0^\infty f(x,\tau) T(\tau,t) d\tau,
 \label{eq:InFS}
\end{equation}
where $f(x,\tau)$ is a continuous approximation to (fluid limit of) $p_n(x)$ and $T(\tau,t)$ is the same kind of approximation to $\chi_n(t)$. Eq. (\ref{eq:InFS}) is called the integral formula of subordination. In what follows we will only work in this limit. 

The parent and the leading processes of the CTRW scheme are processes with independent and identically distributed (i.i.d.) increments, i.e. correspond to sums of i.i.d. random variables. Exactly such sums constitute the elementary (and elemental) class of objects studied in the theory of large deviations. Let $x_n$ be a sequence of independent random variables, $s_n = (x_1+x_2+ ... +x_n)/n$ be the empirical mean of the first $n$ elements in the sequence, and $P_n(s)$ be the probability distribution of $s_n$. Then, under known conditions, there exists the limit
\[
 \mathcal{C}(s) = - \lim_{n \to \infty} \frac{1}{n} \ln P_n(s)
\]
called the rate function (RF), or Cram\'er's function, of large deviations for $s_n$ \cite{Cecconi,Varadhan,Touchette}. The distribution of $s_n$ for sufficiently large $n$ is then given, up to a normalization constant, by 
\[
 P_n(s) \sim \exp \left[-n \mathcal{C}(s) \right].
\]
When making the inverse variable transformation from $s_n$ to the sum $S_n = x_1+x_2 + ... + x_n = n s_n$ of the first $n$ random variables, one sees that its PDF takes the form  
\begin{equation}
 p_n(S) \sim \exp \left[-n \mathcal{C}\left( \frac{S}{n}\right) \right],
 \label{eq:LaDF}
\end{equation}
up to an $n$-dependent normalization constant. Eq. (\ref{eq:LaDF}) will be called the \textit{large deviation form} for the PDF of the corresponding sum. To avoid ambiguity, we will denote the rate functions by scriptized letters. 

At the beginning, only two PDFs are known: the one of the single step lengths, and the one of the waiting times. Therefore, in our discussion, we will start from the rate functions of sums of independent and 
identically distributed random variables (step times and step lengths), and derive the properties of all other rate functions from two rate functions: $\mathcal{I}(t)$ and $\mathcal{R}(x)$, the ones for the  leading process and for the parent process (simple RW), which can be readily calculated. 
The large deviation forms for the PDFs of the corresponding sums are then 
\begin{equation}
 f(x,\tau) \sim \exp\left[-\tau \mathcal{R}\left(\frac{x}{\tau}\right) \right],
\label{eq:LaDF_pp}
\end{equation}
and
\begin{equation}
 p(t,\tau) \sim \exp\left[-\tau \mathcal{I}\left(\frac{t}{\tau}\right) \right].
\label{eq:LaDF_tp}
\end{equation}
To easily manipulate the variables we have to keep the arguments of all non-power-law functions dimensionless. 
To do so, we introduce the natural units of length and time. As a natural unit of length one may choose the mean squared displacement in a single step, and as a natural unit of time a characteristic time of a single step. 
If the mean waiting time for a step exists, we will choose this mean waiting time as the characteristic time. In the cases when the mean waiting time diverges, for the purpose of comparison of different situations, one should choose a characteristic 
time such that the mean number of steps $\langle n(t) \rangle$ made up to time $t$ is the same for all cases compared. Note that this requirement is fulfilled when choosing the mean waiting time, provided it exists, as a characteristic time, as above. Thus, from here on, all our quantities are dimensionless. 
The units can be easily restored in the final expressions when necessary. 

The two derived rate functions, the ones for the directing process and the one for the whole CTRW (a function we seek to know) will be denoted by $\mathcal{T}(\tau)$ and $\mathcal{X}(x)$, respectively. 
The large deviation forms for the PDFs of $\tau$ and $x$ are:  
\begin{equation}
T(\tau,t) \sim \exp \left[-t \mathcal{T}\left( \frac{\tau}{t}\right) \right],
\label{eq:LaDF_dp}
\end{equation}
 and
 \begin{equation}
  p(x,t) \sim \exp \left[-t \mathcal{X}\left( \frac{x}{t}\right) \right].
  \label{eq:LaDF_CTRW}
 \end{equation}

Before starting our derivation, let us go through some important properties of the rate functions \cite{Cecconi,Varadhan,Touchette}. The rate functions for the mean of i.i.d. random variables are convex \cite{Varadhan}, see also 
Appendix \ref{ap:properties}. Cram\'er's theorem states that the rate function $\mathcal{C}(x)$ is the convex conjugate (a Legendre-Fenchel transformation) of the cumulant generation function $\mathcal{L}(q) = \ln \langle e^{xq} \rangle$ of the distribution of $x$ (a simple explanation is given in Ref. \cite{Cecconi}). 
For the sake of brevity we will call this transformation simply a Legendre transformation, as it is done, say, when transforming a Lagrange to a Hamilton function in mechanics. For 
strictly convex rate functions (i.e. for all cases discussed in the present work) this transformation is invertible. The inverse of a Legendre transformation is given by a Legendre transformation again (an involution property). 
Two other properties of rate functions, which are important for what follows, are that the rate function is non-negative, and that it vanishes identically at the value of $x = \mu$ corresponding to the mean of the distribution of the single variable. The rate function can be defined on the whole real line (e.g. for a Gaussian distribution of $x$), on a half-line (like it is for waiting times), or on a finite interval. Outside of its domain of definition (i.e. outside of the domain available for the initial random variables) it is taken to be ``literally infinite'', so that the probability to have such values (the ones outside the domain of definition) is strictly zero.  

To keep equations concise we will introduce the following nonstandard notation for the Legendre transformation: Let $g(x)$ be a convex function defined on the corresponding interval, 
and let a function $f(z)$ of variable $z$ be its Legendre transformation defined as 
\[
 g(z) = \mathrm{sup}_x \left\{zx - f(x)\right\}.
\]
The Legendre transformation is an operator which makes a mapping $f(x) \mapsto g(z)$. Our notation will keep track of the variable's names in the corresponding functions
\[
 g(z) = \widehat{\mathfrak{L}}_{\{z \leftarrow x\}} f(x).
\]
The direction of the arrow is connected with the order of the symbols: the operator $\widehat{\mathfrak{L}}$ acts on the function of $x$ standing right from it, and transforms the function $f$ of variable $x$ into a function $g$ of variable $z$ (the names of the variables are changed from right to the left). Since we choose to denote the variable's names in the notation for the operator, we don't have to put them in the functions at all (but we will).

The main result of the present article is as follows: 
\begin{equation}
 \mathcal{X}\left( z \right) = - \sup_\xi \left[- \frac{ \mathcal{R}( z \xi ) + \mathcal{I}(\xi)}{\xi} \right],
 \label{eq:MainR}
\end{equation}
with $\mathcal{I}(z) = \widehat{\mathfrak{L}}_{\{z \leftarrow q\}} \mathcal{L}(q)$ and $\mathcal{R}(z) = \widehat{\mathfrak{L}}_{\{z \leftarrow q\}} \mathcal{D}(q)$ where $\mathcal{L}(q) = \langle e^{qt }\rangle = \int_{0}^\infty e^{qt} \psi(t) dt$ with $\psi(t)$ being the PDF of waiting times (waiting time density, WTD), and $\mathcal{D}(q) = \langle e^{qx }\rangle = \int_{-\infty}^\infty e^{qx} \lambda(x) dx$ with $\lambda(x)$ being the PDF of displacements in a single step (steps' lengths density, SLD). For a Gaussian distribution of single steps' lengths discussed in \cite{Barkai,Wang}, the result can be put in a closed form involving only the Legendre transformations
\begin{equation}
 \mathcal{X}(x) = - \widehat{\mathfrak{L}}_{\{- x^2/2 \leftarrow \xi\}} \left[ \xi^{-1} \widehat{\mathfrak{L}}_{\{\xi \leftarrow q\}} \mathcal{L}(q) \right].
 \label{eq:CTRWG}
\end{equation}

\section{The large deviations rate for CTRW \label{sec:Main}}

Let us now derive the forms, Eq. (\ref{eq:MainR}) and (\ref{eq:CTRWG}). Substituting the large deviation forms for PDFs of the parent and directing  processes (Eqs.(\ref{eq:LaDF_pp}) and (\ref{eq:LaDF_dp}), respectively) into the integral formula of subordination, Eq. (\ref{eq:InFS}), one gets
\[
 p(x,t) \sim \int_0^\infty \exp \left[-\tau \mathcal{R}\left( \frac{x}{\tau}\right) - t \mathcal{T}\left(\frac{\tau}{t}\right) \right] d \tau.
\]
On the l.h.s., we introduce the large deviation form of the whole CTRW (Eq.(\ref{eq:LaDF_CTRW}))
\[
 \exp\left[-t \mathcal{X}\left( \frac{x}{t}\right)\right] \sim \int_0^\infty \exp \left[-\tau \mathcal{R}\left( \frac{x}{\tau}\right) - t \mathcal{T}\left(\frac{\tau}{t}\right) \right] d \tau
\]
and change to the new variables $z = x/t$ and $\xi = t/\tau$:
\[
 \exp\left[-t \mathcal{X}\left( z \right)\right] \sim 
  \int_0^\infty \frac{t}{\xi^2} \exp \left\{ - t \left[\frac{\mathcal{R}\left( z \xi \right)}{\xi} + \mathcal{T}\left(\frac{1}{\xi}\right) \right] \right\} d \xi.
\]
Now, we use the relation between the RF for the leading and directing processes, which is given by
\begin{equation}
 \mathcal{T}(\tau) = \tau \mathcal{I}\left(\frac{1}{\tau}\right),
 \label{eq:dirlead}
\end{equation}
see Ref. \cite{Glynn}. The proof of this relation (and the discussion of conditions under which it holds) implies a longer chain of mathematical discussions. 
Although this relation is known, we sketch a simple explanation on the physical level of rigor is given in Appendix \ref{ap:conexion}. 

Using this relation one gets
\[
	\exp\left[-t \mathcal{X}\left( z \right)\right] \sim 
	\int_0^\infty \frac{t}{\xi^2} \exp \left\{ - t \left[ \frac{\mathcal{R}\left( z \xi \right)}{\xi} + \frac{\mathcal{I}(\xi)}{\xi} \right] \right\} d \xi.
\]
This integral can then be solved by the Laplace's method \cite{Butler}. Thus, we can write
\[
\exp\left[-t \mathcal{X}\left( z \right)\right] \sim  \exp \left\{ - t \inf_\xi \left[ \frac{\mathcal{R}\left( z \xi \right)}{\xi} + \frac{\mathcal{I}(\xi)}{\xi} \right] \right\},
\]
where the pre-exponential term is disregarded, since it does not contribute to the large deviations when taking the limit $\lim_{t \to \infty} \ln [p(x,t)] / t$. Equating the arguments of the exponentials, we get
\[
 \mathcal{X}\left( z \right) = - \sup_\xi \left[-\frac{\mathcal{R}( z \xi ) + \mathcal{I}(\xi)}{\xi} \right],
\]
which is our main result, Eq. (\ref{eq:MainR}). Note that, the infimum was changed for the supremum of the negative of the expression, which will be useful to later relate this quantity to a Legendre transformation in the case of a Gaussian SLD. Since the large deviation rates $\mathcal{R}(x)$ and $\mathcal{I}(t)$ are readily given by the Legendre transformations of the corresponding cumulant generating functions, the corresponding supremum can be easily calculated for the cases of interest. In what follows, we consider in some detail the case brought to our attention by the works \cite{Barkai,Wang}, namely, the case of Gaussian distribution of single step lengths. For this case, the RF $\mathcal{X}\left( z \right)$ follows from the cumulant generating function of waiting times by two Legendre transformations (see below). However, first, as an example, let us analyse the simplest random walk, i.e., the Bernoulli one.

\subsection{CTRW with a fixed step length}

As a first example, let us consider the Bernoulli random walk, with steps of fixed length $\pm1$ taken with probability $1/2$, respectively,  and the simplest possible leading process, a Poisson process with rate 1, i.e. with $\psi(t) = e^{-t}$. The Bernoulli random walk is essentially the first example of Refs. \cite{Cecconi} and \cite{Varadhan}. For this case 
\[
 p_N(S) = 2^{-N} \frac{N!}{\left(\frac{N}{2}- \frac{S}{2}\right)!\left(\frac{N}{2}+ \frac{S}{2}\right)!}.
\]
Applying Stirling formula we get in the first order in $N$ 
\[
\ln p_N(s) = - \frac{N}{2}\left[(1-s) \ln(1-s) + (1+s) \ln(1+s)\right]
\]
with $s= S/N$, so that 
\[
 \mathcal{R}(x) = \frac{1}{2}\left[(1-x) \ln(1-x) + (1+x) \ln(1+x)  \right].
\]
This function is parabolic around 0 and diverges for $x \to \pm 1$ (and is ``literally infinite'' outside of the interval $(-1,1)$). The function  $\mathcal{I}(\xi) = \widehat{\mathfrak{L}}_{\{\xi \leftarrow q\}} \mathcal{L}(q)$ corresponding to a Poisson process can be readily calculated (and essentially is known since long ago), and reads
\begin{equation}
 \mathcal{I}(\xi) = \xi - 1- \ln \xi.
 \label{eq:Poiss}
\end{equation}
The supremum in Eq.(\ref{eq:MainR}), for fixed $z$ is achieved at $\xi= 1/\sqrt{1+z^2}$, and $\mathcal{X}(z)$ is readily evaluated:
\begin{widetext}
\begin{multline}
	\mathcal{X}(z) =  1 + \sqrt{1+z^2} \left\{ -1+ \ln \sqrt{1+z^2} \right. \\
	 + \left. \frac{1}{2}\left[ \left(1- \frac{z}{\sqrt{1+z^2}} \right) \ln\left(1- \frac{z}{\sqrt{1+z^2}} \right)+  \left(1+ \frac{z}{\sqrt{1+z^2}} \right)\ln\left(1+ \frac{z}{\sqrt{1+z^2}} \right) \right] \right\},
\label{eq:BP}
\end{multline}
\end{widetext}
which is the function with a quadratic behavior close to $z = 0$, $\mathcal{X}(z) \simeq z^2/2 + O(z^4)$, and with the large-$z$ asymptotics being $\mathcal{X}(z) \simeq |z|(1 - \ln2 +\ln|z|)$
(note that the function is not ``cut'' at a finite value of $z$, at difference to the large deviation rate of the parent process). 

\section{CTRW with Gaussian distribution of steps lengths \label{sec:Gauss}}

For the Gaussian distribution of step lengths the large deviations rate for the parent process is given by 
\[
 \mathcal{R}(x) = \frac{x^2}{2}. 
\]
Introducing this result into Eq. (\ref{eq:MainR}), one obtains 
\begin{equation}
 \mathcal{X}\left( x \right) = - \sup_\xi \left[- \frac{x^2 \xi}{2} -  \frac{\mathcal{I}(\xi)}{\xi} \right], 
 \label{eq:CTRW_gauss}
\end{equation}
which is given by a Legendre transformation of the function $\mathcal{I}(\xi)/\xi$ taken at the value of the  Legendre variable $z = - x^2/2$. Using our notation for the Legendre transformations and the fact that  $\mathcal{I}(\xi) = \widehat{\mathfrak{L}}_{\{\xi \leftarrow q\}} \mathcal{L}(q)$ we arrive at the final result in a closed form
\[
 \mathcal{X}(x) = - \widehat{\mathfrak{L}}_{\{- x^2/2 \leftarrow \xi\}} \left[ \xi^{-1} \widehat{\mathfrak{L}}_{\{\xi \leftarrow q\}} \mathcal{L}(q) \right],
\]
Eq. (\ref{eq:CTRWG}). Now we use this formula for discussing several important examples and then make some conclusions about the behavior of the large deviation rate functions of CTRW with Gaussian step lengths' distribution for very large deviations ($x \gg 1$).

\subsection{Exponential waiting times}
\begin{figure}[tbp]
\centering
\def\svgwidth{\columnwidth}
\begingroup%
  \makeatletter%
  \providecommand\color[2][]{%
    \errmessage{(Inkscape) Color is used for the text in Inkscape, but the package 'color.sty' is not loaded}%
    \renewcommand\color[2][]{}%
  }%
  \providecommand\transparent[1]{%
    \errmessage{(Inkscape) Transparency is used (non-zero) for the text in Inkscape, but the package 'transparent.sty' is not loaded}%
    \renewcommand\transparent[1]{}%
  }%
  \providecommand\rotatebox[2]{#2}%
  \newcommand*\fsize{\dimexpr\f@size pt\relax}%
  \newcommand*\lineheight[1]{\fontsize{\fsize}{#1\fsize}\selectfont}%
  \ifx\svgwidth\undefined%
    \setlength{\unitlength}{612bp}%
    \ifx\svgscale\undefined%
      \relax%
    \else%
      \setlength{\unitlength}{\unitlength * \real{\svgscale}}%
    \fi%
  \else%
    \setlength{\unitlength}{\svgwidth}%
  \fi%
  \global\let\svgwidth\undefined%
  \global\let\svgscale\undefined%
  \makeatother%
  \begin{picture}(1,0.76470588)%
    \lineheight{1}%
    \setlength\tabcolsep{0pt}%
    \put(0,0){\includegraphics[width=\unitlength,page=1]{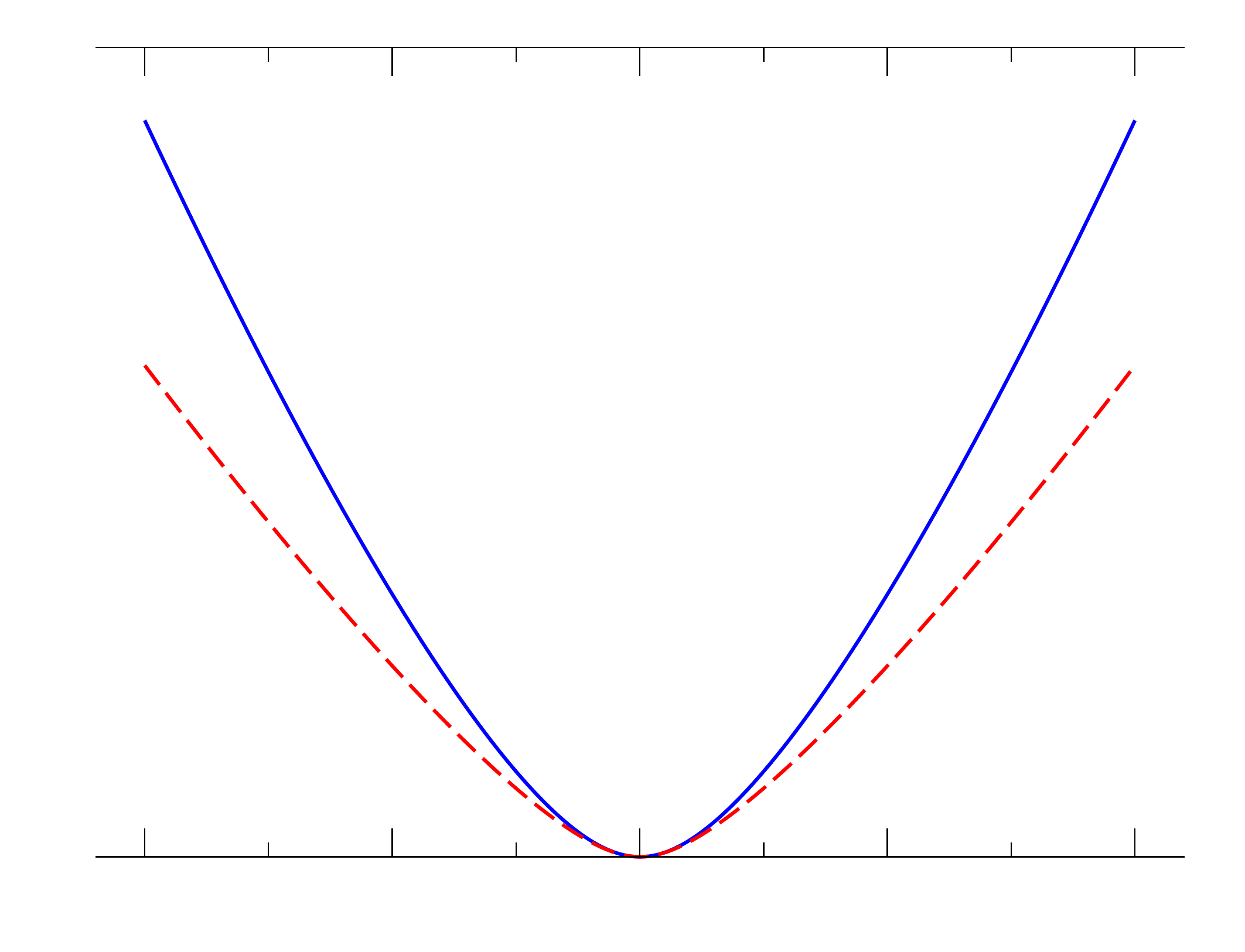}}%
    \put(0.10235049,0.04411765){\makebox(0,0)[lt]{\lineheight{1.25}\smash{\begin{tabular}[t]{l}-10\end{tabular}}}}%
    \put(0.30691356,0.04411765){\makebox(0,0)[lt]{\lineheight{1.25}\smash{\begin{tabular}[t]{l}-5\end{tabular}}}}%
    \put(0.50902549,0.04411765){\makebox(0,0)[lt]{\lineheight{1.25}\smash{\begin{tabular}[t]{l}0\end{tabular}}}}%
    \put(0.70786961,0.04411765){\makebox(0,0)[lt]{\lineheight{1.25}\smash{\begin{tabular}[t]{l}5\end{tabular}}}}%
    \put(0.90099461,0.04411765){\makebox(0,0)[lt]{\lineheight{1.25}\smash{\begin{tabular}[t]{l}10\end{tabular}}}}%
    \put(0.56902549,0.03666667){\makebox(0,0)[lt]{\lineheight{1.25}\smash{\begin{tabular}[t]{l}$x$\end{tabular}}}}%
    \put(0,0){\includegraphics[width=\unitlength,page=2]{Plot1.pdf}}%
    \put(0.05901961,0.06911765){\makebox(0,0)[lt]{\lineheight{1.25}\smash{\begin{tabular}[t]{l}0\end{tabular}}}}%
    \put(0.04901961,0.21042206){\makebox(0,0)[lt]{\lineheight{1.25}\smash{\begin{tabular}[t]{l}5\end{tabular}}}}%
    \put(0.0275817,0.35172631){\makebox(0,0)[lt]{\lineheight{1.25}\smash{\begin{tabular}[t]{l}10\end{tabular}}}}%
    \put(0.0275817,0.49303072){\makebox(0,0)[lt]{\lineheight{1.25}\smash{\begin{tabular}[t]{l}15\end{tabular}}}}%
    \put(0.0275817,0.63433497){\makebox(0,0)[lt]{\lineheight{1.25}\smash{\begin{tabular}[t]{l}20\end{tabular}}}}%
    \put(-0.01522876,0.42411765){\makebox(0,0)[lt]{\lineheight{1.25}\smash{\begin{tabular}[t]{l}$\mathcal{X}(x)$\end{tabular}}}}%
    \put(0,0){\includegraphics[width=\unitlength,page=3]{Plot1.pdf}}%
    \put(0.68202206,0.65169935){\makebox(0,0)[lt]{\lineheight{1.25}\smash{\begin{tabular}[t]{l}\scriptsize Bernoulli\end{tabular}}}}%
    \put(0,0){\includegraphics[width=\unitlength,page=4]{Plot1.pdf}}%
    \put(0.68202206,0.6151634){\makebox(0,0)[lt]{\lineheight{1.25}\smash{\begin{tabular}[t]{l}\scriptsize Gaussian\end{tabular}}}}%
    \put(0,0){\includegraphics[width=\unitlength,page=5]{Plot1.pdf}}%
  \end{picture}%
\endgroup%
\caption{A comparison of the rate functions $\mathcal{X}(x)$ for a Bernoulli CTRW and CTRW with Gaussian distribution of step lengths, both with exponential WTD, Eq. (\ref{eq:BP}). 
For $x \ll 1$, both curves have the same parabolic behavior, which is a consequence of the CLT, see the text for details.
\label{fig:1}
} 
\end{figure}

Let us assume that the waiting times follow a Poisson process, with $\psi(t)=e^{-t}$ and a rate function given by Eq. (\ref{eq:Poiss}).
Now, we define the function
\[
g(\xi) = \frac{\mathcal{I}(\xi)}{\xi} = 1 - \frac{1}{\xi} - \frac{\ln \xi}{\xi},
\]
and perform the Legendre transformation
\[
z = \frac{d}{d \xi} g(\xi) = \frac{\ln \xi}{\xi^2} 
 \] so that
 \[
\xi = \exp \biggr[ - \frac{1}{2} W_0(-2 z) \biggr],
\]
where $W_0(\cdot)$ is the Lambert function. Then, according to Eq. (\ref{eq:CTRWG}), we change the variable to $z = - x^2 / 2$ and invert the sign:
\begin{multline}
\mathcal{X}(x) = 1 - \exp \biggr[ \frac{1}{2} W_0(x^2) \biggr] + \frac{x^2}{2} \exp \biggr[ - \frac{1}{2} W_0(x^2) \biggr] \\
+ \frac{1}{2} W_0(x^2) \exp \biggr[ \frac{1}{2} W_0(x^2) \biggr] .
\label{eq:GP}
\end{multline}
Using the properties of the Lambert function, the two limits, of small and of large $x$, can be found. For $x>>1$, $W_0(x^2) \sim \ln x^2 + \ln \ln x^2$, and $\mathcal{X}(x) \sim |x| \sqrt{2 \ln |x|}$. On the other hand, for $x<<1$, $W_0(x^2) \sim x^2$, and performing a Taylor expansion around zero, one has $\mathcal{X}(x) \sim x^2/2 + \mathcal{O} (x^4)$ corresponding to a Gaussian. Fig. \ref{fig:1} shows a comparison between the rate functions for the Bernoulli case with exponential WTD, 
Eq. (\ref{eq:BP}), and for the Gaussian SLD with exponential WTD, Eq. (\ref{eq:GP}). One can see that, as $x \to 0$, both curves coincide and show a parabolic behavior which is a consequence of the Central Limit Theorem (CLT). For $x>1$ the curves diverge, with the one for the Bernoulli RW growing faster. This shows that the large deviation rate of CTRWs is sensitive to the single step length distribution (and can be used as a probe for such). This discussion reproduces
the results of Ref. \cite{Wang}.

\subsection{One-sided L\'evy law}

Now we take $\psi(t)$ to follow a one-sided L\'evy law with exponent $\alpha$. Its Laplace characteristic function $\widetilde{\psi}(s) = \langle e^{-st} \rangle$ reads: $\widetilde{\psi}(s) = \int_0^\infty \psi(t) e^{-st} dt = \exp(- \sigma s^\alpha)$ 
with $0<\alpha < 1$ (for $s>0$) and 
$\sigma$ being the scale parameter. To allow for further comparison with the behavior for the Pareto-distributed waiting times, we fix this to be $\sigma =1/\Gamma(\alpha+1)$. This fixes the characteristic time of a step. 
The mean number of steps $\langle n(t) \rangle$ done during the time $t$ is then 
\begin{equation}
 \langle n(t) \rangle = t^{\alpha}.
 \label{eq:noft}
\end{equation}
Let us denote $C_{L} = \Gamma(\alpha+1)^{1/\alpha}$.
Thus, $\widetilde{\psi}(s) = \exp[- (s / C_L)^{\alpha}]$, so that $f(q)= \langle e^{qt}\rangle =  \widetilde{\psi}(-q)$ is given by
\[
 f(q) = \left\{ \begin{array}{cl}
                 \exp[- (-q)^\alpha / C_L^{\alpha}] & q \leq 0 \\
                 +\infty & q > 0
                \end{array}
\right. 
\]
and
\[
 \mathcal{L}(q) =  \left\{ \begin{array}{cl}
                 - (-q)^\alpha / C_L^{\alpha} & q \leq 0 \\
                 +\infty & q > 0
                \end{array}
\right. .
\]
Now, we perform the Legendre transform:
\[
 t = \frac{d}{dq}\mathcal{L}(q) = \frac{\alpha (-q)^{\alpha -1}}{C_L^{\alpha}}
 \longrightarrow q = - \biggr(\frac{C_L^{\alpha} t}{\alpha}\biggr)^{\frac{1}{\alpha-1}},
\]
which leads to 
\[
 \mathcal{I}(t) = (1-\alpha)\left(\frac{C_L t}{\alpha} \right)^{-\frac{\alpha}{1-\alpha}}.
\]
Now 
\[
g(\xi) = \frac{\mathcal{I}(\xi)}{\xi} = (1-\alpha)\biggr(\frac{\alpha}{C_L}\biggr)^{\frac{\alpha}{1-\alpha}} \xi^{-\frac{1}{1-\alpha}},
\]
and now the second Legendre transformation can be performed:
\[
 z = \frac{d}{d\xi} g(\xi) = - \biggr(\frac{\alpha}{C_L}\biggr)^{\frac{\alpha}{1-\alpha}} \xi^{-\frac{2-\alpha}{1-\alpha}},
\]
with 
\[
 \xi = \biggr( \frac{C_L}{\alpha} \biggr)^{-\frac{\alpha}{2-\alpha}}\left(-z\right)^{-\frac{1-\alpha}{2-\alpha}}.
\]
Finally, the rate function of displacements reads 
\begin{equation}
\mathcal{X}(x)= (2-\alpha) 2^{-\frac{1}{2-\alpha}} \biggr(\frac{\alpha}{C_L}\biggr)^{\frac{\alpha}{2-\alpha}}  |x|^{\frac{2}{2-\alpha}}. 
\label{eq:Levy}
\end{equation}
For $\alpha \to 1$ it tends to a parabolic RF of a Gaussian, for $\alpha \to 0$ to a wedge-like RF of ultraslow diffusion with a purely exponential PDF tails, see \cite{KlaSok}.

\subsection{Pareto distributions}

Now let us consider two other WTDs, namely the Pareto distributions of types I and II with parameter $\alpha$ (for the sake of brevity these will be sometimes referred to as
Pareto I and Pareto II distributions). Both distributions have the same asymptotic behavior for long times, $\psi(t) \propto t^{-1-\alpha}$ but 
differ with respect to their short time behavior. The distributions with $0< \alpha \leq 1$ are fat-tailed and do not possess the mean, the distributions with $\alpha > 1$ do possess the mean waiting time. 

To be able to compare them to each other, and to a L\'evy distribution for $\alpha \in (0,1)$, one has to fix the time scale. For $\alpha>1$, both Pareto distributions possess a mean waiting time, so the scaling factor is this last quantity. 
For $0<\alpha \leq 1$, we scale the distributions in such a way that  the mean number of steps $\langle n(t)\rangle$ for a given $t$ is the same for both distributions and exactly the same as for the L\'evy case (Eq. (\ref{eq:noft})).
Thus, the Pareto type I WTD $\psi(t)$ is defined as 
\begin{equation}
  \psi(t) = \left\{
     \begin{array}{cc}
        0 & 0 \geq t > C_{I} \\\\
        \displaystyle \frac{\alpha C_{I}^{\alpha}}{ t^{\alpha+1}} & t\geq C_{I}
        
     \end{array}
   \right.
\label{eq:paretoI}
\end{equation}
where $C_{I} = [\Gamma(1+\alpha)\Gamma(1-\alpha)]^{-1/\alpha}$ if $\alpha \leq 1$, or $C_{I} = (\alpha - 1)/\alpha$ if $\alpha>1$. Its Laplace transform reads
\begin{equation}
\widetilde{\psi}(s) = \alpha C_{I}^{\alpha} s^{\alpha} \Gamma(-\alpha,C_{I} s),
\label{eq:paretoI_lap}
\end{equation}
with $\Gamma(x,y)$ being an upper incomplete $\Gamma$-function. On the other hand, the Pareto type II WTD can be defined as
\begin{equation}
\psi(t) = \frac{\alpha C_{II}^{\alpha}}{(C_{II}+t)^{\alpha+1}},
\label{eq:paretoII}
\end{equation}
where $C_{II} = [\Gamma(1+\alpha)\Gamma(1-\alpha)]^{-1/\alpha}$ if $\alpha \leq 1$, and $C_{II} = (\alpha - 1)$ if $\alpha>1$. Its Laplace transform reads
\begin{equation}
\widetilde{\psi}(s) = \alpha C_{II}^{\alpha} s^{\alpha} e^{C_{II} s} \Gamma(-\alpha,C_{II} s).
\label{eq:paretoII_lap}
\end{equation}
Due to the presence of incomplete Gamma functions, the Legendre transforms have to be performed numerically, by solving algebraic equations. 

\begin{figure}[h!]
\centering
\def\svgwidth{\columnwidth}
\begingroup%
  \makeatletter%
  \providecommand\color[2][]{%
    \errmessage{(Inkscape) Color is used for the text in Inkscape, but the package 'color.sty' is not loaded}%
    \renewcommand\color[2][]{}%
  }%
  \providecommand\transparent[1]{%
    \errmessage{(Inkscape) Transparency is used (non-zero) for the text in Inkscape, but the package 'transparent.sty' is not loaded}%
    \renewcommand\transparent[1]{}%
  }%
  \providecommand\rotatebox[2]{#2}%
  \newcommand*\fsize{\dimexpr\f@size pt\relax}%
  \newcommand*\lineheight[1]{\fontsize{\fsize}{#1\fsize}\selectfont}%
  \ifx\svgwidth\undefined%
    \setlength{\unitlength}{612bp}%
    \ifx\svgscale\undefined%
      \relax%
    \else%
      \setlength{\unitlength}{\unitlength * \real{\svgscale}}%
    \fi%
  \else%
    \setlength{\unitlength}{\svgwidth}%
  \fi%
  \global\let\svgwidth\undefined%
  \global\let\svgscale\undefined%
  \makeatother%
  \begin{picture}(1,0.76470588)%
    \lineheight{1}%
    \setlength\tabcolsep{0pt}%
    \put(0,0){\includegraphics[width=\unitlength,page=1]{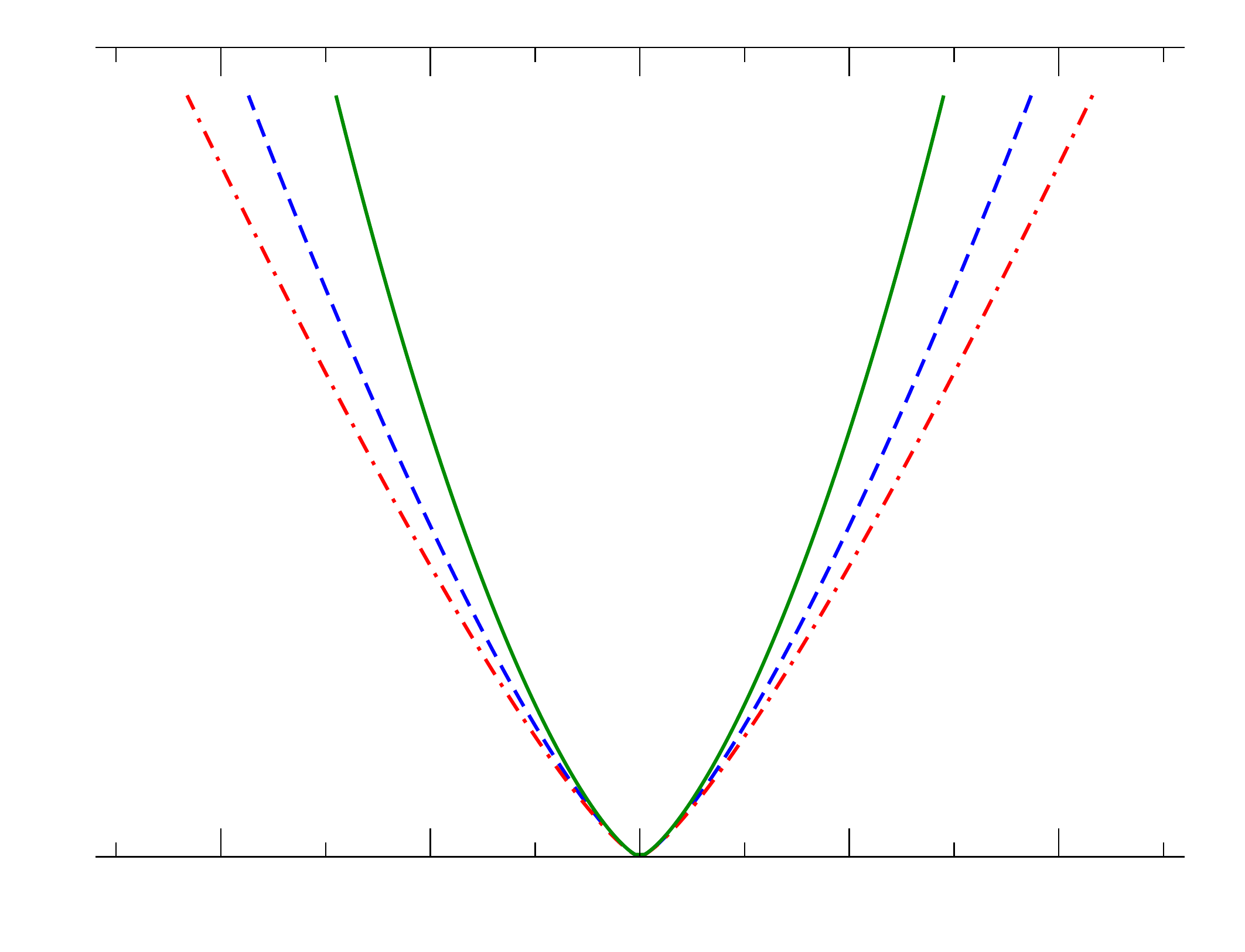}}%
    \put(0.16353333,0.04411765){\makebox(0,0)[lt]{\lineheight{1.25}\smash{\begin{tabular}[t]{l}-10\end{tabular}}}}%
    \put(0.3375049,0.04411765){\makebox(0,0)[lt]{\lineheight{1.25}\smash{\begin{tabular}[t]{l}-5\end{tabular}}}}%
    \put(0.50902549,0.04411765){\makebox(0,0)[lt]{\lineheight{1.25}\smash{\begin{tabular}[t]{l}0\end{tabular}}}}%
    \put(0.6772781,0.04411765){\makebox(0,0)[lt]{\lineheight{1.25}\smash{\begin{tabular}[t]{l}5\end{tabular}}}}%
    \put(0.83981193,0.04411765){\makebox(0,0)[lt]{\lineheight{1.25}\smash{\begin{tabular}[t]{l}10\end{tabular}}}}%
    \put(0.57902549,0.01666667){\makebox(0,0)[lt]{\lineheight{1.25}\smash{\begin{tabular}[t]{l}$x$\end{tabular}}}}%
    \put(0,0){\includegraphics[width=\unitlength,page=2]{Plot2.pdf}}%
    \put(0.04901961,0.06911765){\makebox(0,0)[lt]{\lineheight{1.25}\smash{\begin{tabular}[t]{l}0\end{tabular}}}}%
    \put(0.04901961,0.26029412){\makebox(0,0)[lt]{\lineheight{1.25}\smash{\begin{tabular}[t]{l}5\end{tabular}}}}%
    \put(0.0325817,0.45147059){\makebox(0,0)[lt]{\lineheight{1.25}\smash{\begin{tabular}[t]{l}10\end{tabular}}}}%
    \put(0.0325817,0.64264706){\makebox(0,0)[lt]{\lineheight{1.25}\smash{\begin{tabular}[t]{l}15\end{tabular}}}}%
    \put(-0.02522876,0.35411765){\makebox(0,0)[lt]{\lineheight{1.25}\smash{\begin{tabular}[t]{l}$\mathcal{X}(x)$\end{tabular}}}}%
    \put(0,0){\includegraphics[width=\unitlength,page=3]{Plot2.pdf}}%
    \put(0.73319853,0.21143791){\makebox(0,0)[lt]{\lineheight{1.25}\smash{\begin{tabular}[t]{l}\scriptsize L $\alpha=0.5$\end{tabular}}}}%
    \put(0,0){\includegraphics[width=\unitlength,page=4]{Plot2.pdf}}%
    \put(0.73319853,0.17326797){\makebox(0,0)[lt]{\lineheight{1.25}\smash{\begin{tabular}[t]{l}\scriptsize P II $\alpha=0.5$\end{tabular}}}}%
    \put(0,0){\includegraphics[width=\unitlength,page=5]{Plot2.pdf}}%
    \put(0.73319853,0.14){\makebox(0,0)[lt]{\lineheight{1.25}\smash{\begin{tabular}[t]{l}\scriptsize P I $\alpha=0.5$\end{tabular}}}}%
    \put(0,0){\includegraphics[width=\unitlength,page=6]{Plot2.pdf}}%
  \end{picture}%
\endgroup%
\caption{A comparison of the rate functions $\mathcal{X}(x)$ for the CTRW with Gaussian SLD and the WTDs following a one-sided L\'evy law, and Pareto type I and type II distributions, all with  with $\alpha = 0.5$
and the same $\langle n(t) \rangle$ as given by Eq. (\ref{eq:noft}). \label{fig:2}}
\end{figure}

Fig. \ref{fig:2} shows a comparison between the rate functions $\mathcal{X}(x)$ of displacements 
in the CTRW with Gaussian STD and the three following WTDs which do not possess mean waiting time: one-sided L\'evy law with $\alpha = 0.5$, and Pareto I and II distributions, both with $\alpha = 0.5$, all with the same $\langle n(t)\rangle$ as
given by Eq. (\ref{eq:noft}). As in the case of the distributions with finite mean waiting times, the curves coincide in the central domain (although now they are not Gaussian but are given by Eq. (\ref{eq:Levy})), but deviate for $x$ large.

\begin{figure}[h!]
\centering
\def\svgwidth{\columnwidth}
\begingroup%
  \makeatletter%
  \providecommand\color[2][]{%
    \errmessage{(Inkscape) Color is used for the text in Inkscape, but the package 'color.sty' is not loaded}%
    \renewcommand\color[2][]{}%
  }%
  \providecommand\transparent[1]{%
    \errmessage{(Inkscape) Transparency is used (non-zero) for the text in Inkscape, but the package 'transparent.sty' is not loaded}%
    \renewcommand\transparent[1]{}%
  }%
  \providecommand\rotatebox[2]{#2}%
  \newcommand*\fsize{\dimexpr\f@size pt\relax}%
  \newcommand*\lineheight[1]{\fontsize{\fsize}{#1\fsize}\selectfont}%
  \ifx\svgwidth\undefined%
    \setlength{\unitlength}{612bp}%
    \ifx\svgscale\undefined%
      \relax%
    \else%
      \setlength{\unitlength}{\unitlength * \real{\svgscale}}%
    \fi%
  \else%
    \setlength{\unitlength}{\svgwidth}%
  \fi%
  \global\let\svgwidth\undefined%
  \global\let\svgscale\undefined%
  \makeatother%
  \begin{picture}(1,0.76470588)%
    \lineheight{1}%
    \setlength\tabcolsep{0pt}%
    \put(0,0){\includegraphics[width=\unitlength,page=1]{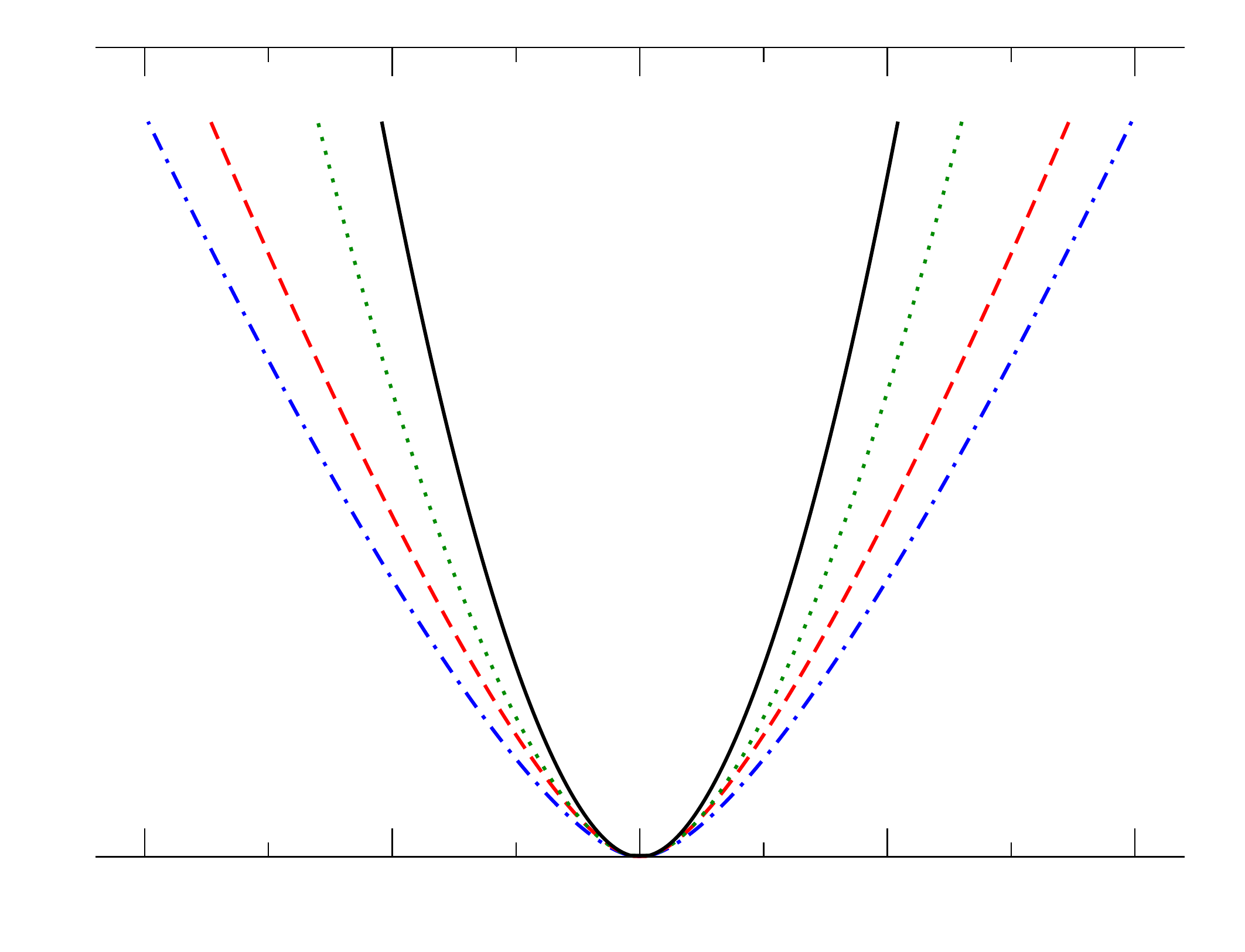}}%
    \put(0.10235049,0.04411765){\makebox(0,0)[lt]{\lineheight{1.25}\smash{\begin{tabular}[t]{l}-10\end{tabular}}}}%
    \put(0.30691356,0.04411765){\makebox(0,0)[lt]{\lineheight{1.25}\smash{\begin{tabular}[t]{l}-5\end{tabular}}}}%
    \put(0.50902549,0.04411765){\makebox(0,0)[lt]{\lineheight{1.25}\smash{\begin{tabular}[t]{l}0\end{tabular}}}}%
    \put(0.70786961,0.04411765){\makebox(0,0)[lt]{\lineheight{1.25}\smash{\begin{tabular}[t]{l}5\end{tabular}}}}%
    \put(0.90099461,0.04411765){\makebox(0,0)[lt]{\lineheight{1.25}\smash{\begin{tabular}[t]{l}10\end{tabular}}}}%
    \put(0.57902549,0.01666667){\makebox(0,0)[lt]{\lineheight{1.25}\smash{\begin{tabular}[t]{l}$x$\end{tabular}}}}%
    \put(0,0){\includegraphics[width=\unitlength,page=2]{Plot3.pdf}}%
    \put(0.04901961,0.06911765){\makebox(0,0)[lt]{\lineheight{1.25}\smash{\begin{tabular}[t]{l}0\end{tabular}}}}%
    \put(0.04901961,0.18729951){\makebox(0,0)[lt]{\lineheight{1.25}\smash{\begin{tabular}[t]{l}2\end{tabular}}}}%
    \put(0.04901961,0.30548121){\makebox(0,0)[lt]{\lineheight{1.25}\smash{\begin{tabular}[t]{l}4\end{tabular}}}}%
    \put(0.04901961,0.42366307){\makebox(0,0)[lt]{\lineheight{1.25}\smash{\begin{tabular}[t]{l}6\end{tabular}}}}%
    \put(0.04901961,0.54184493){\makebox(0,0)[lt]{\lineheight{1.25}\smash{\begin{tabular}[t]{l}8\end{tabular}}}}%
    \put(0.0325817,0.6600268){\makebox(0,0)[lt]{\lineheight{1.25}\smash{\begin{tabular}[t]{l}10\end{tabular}}}}%
    \put(-0.02522876,0.35411765){\makebox(0,0)[lt]{\lineheight{1.25}\smash{\begin{tabular}[t]{l}$\mathcal{X}(x)$\end{tabular}}}}%
    \put(0,0){\includegraphics[width=\unitlength,page=3]{Plot3.pdf}}%
    \put(0.45790441,0.65496732){\makebox(0,0)[lt]{\lineheight{1.25}\smash{\begin{tabular}[t]{l}\scriptsize P II $\alpha=1.5$\end{tabular}}}}%
    \put(0,0){\includegraphics[width=\unitlength,page=4]{Plot3.pdf}}%
    \put(0.45790441,0.62169935){\makebox(0,0)[lt]{\lineheight{1.25}\smash{\begin{tabular}[t]{l}\scriptsize P II $\alpha=2.5$\end{tabular}}}}%
    \put(0,0){\includegraphics[width=\unitlength,page=5]{Plot3.pdf}}%
    \put(0.45790441,0.58843137){\makebox(0,0)[lt]{\lineheight{1.25}\smash{\begin{tabular}[t]{l}\scriptsize P I $\alpha=1.5$\end{tabular}}}}%
    \put(0,0){\includegraphics[width=\unitlength,page=6]{Plot3.pdf}}%
    \put(0.45790441,0.5551634){\makebox(0,0)[lt]{\lineheight{1.25}\smash{\begin{tabular}[t]{l}\scriptsize P I $\alpha=2.5$\end{tabular}}}}%
    \put(0,0){\includegraphics[width=\unitlength,page=7]{Plot3.pdf}}%
  \end{picture}%
\endgroup%
\caption{A comparison of the rate function $\mathcal{X}(x)$ for the CTRW with Gaussian SLD and the following WTDs: Pareto I distributions with $\alpha = 1.5, 2.5$, and Pareto II with $\alpha=1.5, 2.5$. Here, the variable $x$ is the quotient between the position and the time. All these WTDs have a mean waiting time. Note the difference in the asymptotic behavior, which for the case of Pareto I is quadratic (Eq. (\ref{eq:asymptotic_PI})), whereas for the Pareto II is linear with a slowly varying  correction (Eq. (\ref{eq:asymptotic_PII})).
\label{fig:3}}
\end{figure}

Fig. \ref{fig:3} shows the corresponding comparison between the Pareto type I and Pareto type II WTDs for the cases $\alpha = 1.5$ and $\alpha=2.5$ when they do possess the mean waiting time. In this case the behavior for small $x$ is 
universally parabolic, as it should be, but the RFs for Pareto type I WTDs universally grow faster at large $x$ than those for Pareto type II WTDs with the same $\alpha$.

We now discuss in more detail these asymptotic growth properties discussing the domain of very large deviations. The cases of the Pareto distributions give enough physical intuition to understand the general behavior of very large deviations. 

\section{Very large deviations for Pareto waiting time densities}

Physically, the behavior of very large deviations ($x^2 \to \infty$) is dominated by realizations in which the number of steps is unusually large, and thus is governed by the behavior of $\psi(t)$ for very short $t$. 
Therefore the two Pareto cases serve as examples for the cases when $\psi(t)$ tends to a constant limit for $t \to 0$ (Pareto II), and when it vanishes identically for $t \to 0$ (Pareto I). There is no wonder that the L\'evy case with $\psi(0)=0$, but shooting up faster than any power of $t$ for non-vanishing but small $t$, shows the behavior in-between of these two extrema. 

Note that the expressions for both Pareto distributions (Eqs. (\ref{eq:paretoI}) and (\ref{eq:paretoII})) are the same for all values of the parameter $\alpha$, the only difference being the values of the scaling parameters $C_I$ and $C_{II}$. Hence, all subsequent results hold independent of the value of $\alpha$.

\subsection{Pareto Type I WTD}

First, let us consider the Pareto I WTD (Eq. (\ref{eq:paretoI})), and let us work with its Laplace transform as given by Eq. (\ref{eq:paretoI_lap}). The small time behavior ($t \to 0$) of the WTD is mirrored in the asymptotic behavior 
of its Laplace transform for $s \to \infty$, see Appendix \ref{ap:asymptotic} for additional discussions. For the Pareto I PDF this asymptotic behavior is given by
\begin{equation}
\widetilde{\psi}(s) \sim \alpha C_{I}^{-1}s^{-1} e^{-C_{I} s},
\label{eq_LapIas}
\end{equation}
which form can alternatively be obtained either by using the asymptotic expansion of the incomplete Gamma function, or by evaluating the corresponding integral for the Laplace transform using the Laplace method.  
From this form it follows that
\[
\mathcal{L}(q) = \ln \alpha - \ln C_{I}  - \ln(-q) + C_{I} q.
\]
Performing the Legendre transformation we get 
\[
t = \frac{d \mathcal{L}(q)}{d q} = C_{I} - \frac{1}{q} \quad \rightarrow \quad q = \frac{1}{C_{I}-t},
\]
and finally obtain
\[
\mathcal{I}(t) = -\ln(t-C_{I})
\]
in the leading order. Now, 
\[
g(\xi) = \frac{\mathcal{I}(\xi)}{\xi} = -\frac{\ln (\xi-C_{I})}{\xi},
\]
and the second Legendre transform can be  performed:
\[
-\frac{x^2}{2} = \frac{d g(\xi)}{d \xi} =- \frac{1}{\xi(\xi-C_I)} + \frac{\ln(\xi-C_I)}{\xi^2}.
\]
Making the change of variable $u=\xi-C_I$, it can be rewritten as
\[
-\frac{x^2}{2} = -\frac{1}{u(u+C_{I})} + \frac{\ln u}{(u + C_{I})^2}
\]
Very large and negative values of the l.h.s. correspond to $u \to 0$, so that
\[
-\frac{x^2}{2} = -\frac{1}{u C_{I}} + \frac{\ln u}{C_{I}^2} 
\]
To invert this expression, one can apply de Bruijn's Theorem for slowly varying functions, see \cite{SVF}. Hence, one ends up with
\[
u = \frac{2}{C_{I} x^2} \biggr[1 + \frac{\ln (C_{I} x^2)}{C_{I}^2 x^2} \biggr],
\]
and, going back to the variable $\xi$:
\[
\xi = \frac{2}{C_{I} x^2} \biggr[1 + \frac{\ln (C_{I} x^2)}{C_{I}^2 x^2}  \biggr] + C_{I}.
\]
Finally, the asymptotic behavior ($|x| \to + \infty$) of the rate function for the CTRW has the form
\begin{equation}
\mathcal{X}(x) \sim \frac{C_{I}}{2} x^2 + \frac{2}{C_{I}} \ln \lvert x \lvert,
\label{eq:asymptotic_PI}
\end{equation}
which is basically a quadratic behavior with a correction given by a slowly varying function. Note that, apart from the value of the constant $C_I$ being a function of $\alpha$, Eq. (\ref{eq:asymptotic_PI}) does not depends on the parameter $\alpha$.

\subsection{Pareto Type II WTD }
Let us now consider the case of a Pareto II WTD (Eq. (\ref{eq:paretoII})), which in the Laplace domain is given by Eq. (\ref{eq:paretoII_lap}). Following the same procedure as for the Pareto I WTD, let us consider the asymptotic behavior ($s \to \infty$) of Eq. (\ref{eq:paretoII_lap}) given by
\[
\widetilde{\psi}(s) \sim \alpha C_{II}^{-1}s^{-1},
\]
(the difference with Eq. (\ref{eq_LapIas}) is the absence of the exponential cutoff for very large $s$), which allow us to obtain the asymptotics of the cumulant generating function:
\begin{equation}
\mathcal{L}(q) = \ln \alpha - \ln C_{II} - \ln(-q).
\label{eq:LaL}
\end{equation}
Applying the Legendre transformation we get:
\[
t = \frac{d \mathcal{L}(q)}{d q} = - \frac{1}{q} \quad \rightarrow \quad q = -\frac{1}{t},
\]
so that the function $\mathcal{I}(t)$ in the leading order is given by 
\[
\mathcal{I}(t) = -\ln t.
\]
Then we construct
\[
\frac{\mathcal{I}(\xi)}{\xi} = - \frac{\ln \xi}{\xi},
\]
and perform the second Legendre transformation
\[
-\frac{x^2}{2} = \frac{d g(\xi)}{d \xi} = \frac{-1+\ln(\xi)}{\xi^2}.
\]
The values of $|x| \to \infty$ correspond to $\xi \to 0$, so that 
\[
\frac{x^2}{2} = -\frac{\ln \xi}{\xi^2}.
\]
To invert this expression, one  applies again the de Bruijn's Theorem. Hence, the inverse reads
\[
\xi = \lvert x \lvert^{-1} \sqrt{2 \ln \lvert x \lvert}.
\]
Finally, the asymptotic behavior ($|x| \to + \infty$) of the rate function for the CTRW is
\begin{equation}
\mathcal{X}(x) \sim \lvert x \lvert \sqrt{2 \ln \lvert x \lvert},
\label{eq:asymptotic_PII}
\end{equation}
which is an essentially linear behavior, with a correction given by a slowly varying function. This result does not depend on the value of the parameter $\alpha$ at all
(and not only up to the parameter values, like in the Pareto I case). 

\subsection{Erlang distributions}

A similar analysis can be performed for Gamma-distributions (Erlang distributions) discussed in \cite{Barkai,Wang}
\[
\psi(t) =  \frac{\lambda^n t^{n-1} e^{-\lambda t}}{(n-1)!},
\]
with $n \in \{1,2,3,\dots\}$, and $\lambda \in (0,\infty)$. The Laplace characteristic function of the Erlang distributions has the following form
\[
\widetilde{\psi}(s) =  \lambda^n (\lambda + s)^{-n}.
\]
Then the asymptotic form of $\mathcal{L}(q)$ differs from Eq.(\ref{eq:LaL}) only by an additional proportionality factor in front of $\ln(-q)$:
\[
\mathcal{L}(q) = -n \ln (-q),
\]
in the limit $q \to -\infty$. Therefore, the essentially linear behavior 
with a slowly varying correction ensues. The difference between the  L\'evy and Pareto type I cases on one hand and Pareto type II and Erlang cases on the other hand is the fact that for the first class of 
distributions the WTD vanishes at zero together with all its derivatives, while in the second situation this is no longer the case. The full classification of possible behaviors will be discussed elsewhere. 
The lesson learned from these examples is that the essentially linear behavior of the large deviation rate function in CTRW is not universal but is pertinent to the specific classes of the waiting time distributions.
The large deviation behavior of displacements probes the WTD for very short waiting times and is therefore a test for microscopic dynamics of the system.

\section{Summary \label{sec:Sum}}
In this paper, we presented a general procedure to compute the rate function for large deviations of displacements in CTRW in a general setting, i.e. for any step length distribution and any waiting time distribution. 
The situation with Gaussian step length distribution is especially simple. In this case the rate function for displacement is given by a sequence of Legendre transforms of the cumulant generating function for waiting times. 
The general discussion is accompanied by analysing important particular examples like the one-sided L\'evy and the Pareto-distributed waiting times. This discussion shows that the large deviation in displacement probe 
the waiting time density for very short times. The essentially linear behavior of the rate function for very large deviations is specific only for situations in which waiting time density does not vanish too fast when the waiting times approach zero. 

\begin{acknowledgments}
The work of APP was financially supported by ``Doctoral Programmes in Germany'' funded by the Deutscher Akademischer Austauschdienst (DAAD) (Programme ID 57440921).
\end{acknowledgments}

\appendix

\section{Properties of the rate function \label{ap:properties}}
For completeness, we present here a sketch of the proof for some of the properties of the rate function, the ones mentioned in the main text of this work. For a more detailed discussion of the proof, see \cite{Dembo}.  
The rate function $\mathcal{C}(x)$ is defined as
\begin{equation}
\mathcal{C}(x) = \sup_{q} [ q x - \ln f(q) ] 
\label{eq:LDRApp}
\end{equation}
where $f(q) = \langle e^{qx} \rangle$ is the moment generating function (note that we define $\mathcal{L}(q) = \ln f(q)$ in the main text). 

With the definition, Eq.(\ref{eq:LDRApp}), the rate function is a convex non-negative function which satisfies $\mathcal{C}(x = \mu) = 0$. We will prove here this last statement. 
In order to prove the convexity, we simple compute for $\lambda \in (0,1)$
\begin{align*}
\mathcal{C}[\lambda & x_1 + (1-\lambda)x_2] = \\
&= \sup_{q} \{ q [\lambda x_1 + (1-\lambda)x_2] - \ln f(q) \} \\
&= \sup_{q} \{ \lambda [ q x_1 - \ln f(q) ] + (1-\lambda) [ q x_2 - \ln f(q) ] \} \\
&\leq \lambda  \sup_{q} [ q x_1 - \ln f(q) ] + (1-\lambda) \sup_{q} [ q x_2 - \ln f(q) ] \\
&= \lambda \mathcal{C}(x_1) + (1-\lambda) \mathcal{C}(x_2).
\end{align*}
Now, to prove the non-negativity of the rate function, let us use the fact that $f(0) = 1$, so $\ln f(0) = 0$. Now, since $\mathcal{C}(x) \geq qx - \ln f(q)$, taking $q = 0$ implies $ \mathcal{C}(x) \geq 0 x - \ln f(0) = 0$ and the non-negativity is established. Now, let us use Jensen's inequality which can be applied since $f(q)$ is a convex function, then
\[
f(q) = \langle \exp(q x) \rangle \leq \exp( q \langle x \rangle ) = \exp( q \mu).
\]
Next, by applying the natural logarithm in both sides of the inequality, it can be expressed as 
\[
q \mu - \ln f(q) \leq 0.
\]
Finally, because of the non-negativity of the rate function, one concludes 
\[
\mathcal{C}(\mu) = 0 = \min_{x \in \mathbb{R}} \mathcal{C}(x).
\]

\section{Relation between the rate functions for the leading and directing processes\label{ap:conexion}}

The relation between the large deviation rates for the leading and the directing process of the CTRW scheme foots on the known relation between the cumulative distribution functions (CDFs) for the leading and directing processes \cite{Baule,Gorenflo1}.
Let  $p(\tau|t) = T(\tau,t)$ be probability density to have exactly $\tau$ steps up to time $t$ (i.e. the probability density of $\tau$ conditioned on $t$), and  $p(t | \tau)$ the corresponding density of the time of the 
last step conditioned on the number of steps. Then, from the fact that $t$ is monotonically non-decreasing function of $\tau$, the integral relation between the two follows. In the continuous limit this relation takes the form
\begin{equation}
 \int_{\tau}^\infty p(\tau' | t) d \tau' = \int_0^t p(t' | \tau) dt'.
 \label{eq:Integral2}
\end{equation}
We note that the rate function $\mathcal{I}(z)$ is already known. From the general properties of the rate functions it follows that $\mathcal{I}(z)$ is a convex function, monotonically non-growing
(in our case, essentially monotonically decaying) for $z < z_0$ with $z_0$ being equal to the mean waiting time, and monotonically non-decaying (in our case growing) 
for $z > z_0$. If the mean diverges, the function is always monotonically decaying. 
From Eq. (\ref{eq:Integral2}) it follows that
\[
 p(\tau | t) = - \frac{d}{d \tau} \int_0^t p(t' | \tau) d t',
\]
or, equivalently,
\[
 p(\tau | t) = \frac{d}{d \tau}\left[1- \int_0^t p(t' | \tau) d t'\right] = \frac{d}{d \tau} \int_t^\infty p(t' | \tau) d t'.
 \]
Substituting the large deviation forms we get
\begin{widetext}
\begin{equation}
\exp\left[- t \mathcal{T} \left(\frac{\tau}{t} \right) \right] \sim - \frac{d}{d \tau} \int_0^t \exp\left[- \tau \mathcal{I} \left(\frac{t'}{\tau} \right) \right] d t' = \int_0^t \left[\mathcal{I} \left(\frac{t'}{\tau}\right) - \frac{t'}{\tau} \mathcal{I}' \left(\frac{t'}{\tau}\right) \right] \exp\left[- \tau \mathcal{I} \left(\frac{t'}{\tau} \right) \right] d t' 
\label{eq:less}
\end{equation}
or
\begin{equation}
\exp\left[- t \mathcal{T} \left(\frac{\tau}{t} \right) \right] \sim  \frac{d}{d \tau} \int_t^\infty \exp\left[- \tau \mathcal{I} \left(\frac{t'}{\tau} \right) \right] d t' = - \int_t^\infty \left[\mathcal{I} \left(\frac{t'}{\tau}\right) - \frac{t'}{\tau} \mathcal{I}' \left(\frac{t'}{\tau}\right) \right] \exp\left[- \tau \mathcal{I} \left(\frac{t'}{\tau} \right) \right] d t'  
\label{eq:more}
\end{equation}
\end{widetext}
The expression in the non-exponential part of both Eqs. (\ref{eq:less}) and (\ref{eq:more}),
\[
 f(x) = \mathcal{I} (x) - x \mathcal{I}'(x)
\]
(with $x = t'/\tau$) is nonnegative for $x < z_0$ and non-positive for $x > z_0$. The first statement (for $x \geq 0$) follows immediately from
the fact that $\mathcal{I} (x)$ is nonnegative and its derivative for $x < z_0$ non-positive. The second statement is slightly finer and
follows from the the relation 
\begin{equation}
 g(z) \geq g(y) + g'(y)(z-y) 
 \label{mainIneq}
\end{equation}
for convex differentiable functions, which we rewrite as
\[
 g(y) - y g'(y) \leq g(z) - g'(y) z.
\]
Now one takes $g(y) = \mathcal{I}(y)$ and $z=z_0$, so that $g(z_0) = \mathcal{I}(z_0)$ vanishes, and notes that for $y > z_0$ the derivative $g'(y) = \mathcal{I}'(y)$ is nonnegative. Therefore the prefactors of the exponentials
in both integrals on the r.h.s. of Eqs. (\ref{eq:less}) and (\ref{eq:more}) are non-negative (essentially, positive for non-degenerated cases). 
For $z < z_0$ the function $\mathcal{I} \left(z \right)$ is monotonically decaying ($\mathcal{I}' (z) <0$), and the argument of the exponential is therefore monotonically growing towards the upper integration boundary. 
For $z > z_0$ the function $\mathcal{I} \left(z \right)$ is monotonically growing ($\mathcal{I}' (z) > 0$), and the absolute maximum of the integrand is achieved on the lower integration boundary. 

To get the relation between $\mathcal{I}$ and $\mathcal{T}$, the argument of the function on the l.h.s. has to be kept constant. Fixing $x= t/\tau$ and changing the variable of integration on the r.h.s. to $x' = t'/\tau$ one obtains:
\[
 \exp\left[- \tau x \mathcal{T} \left(\frac{1}{x} \right) \right] \sim \tau \int_a^b |f(x')|  \exp\left[- \tau \mathcal{I} \left(x'\right) \right] dx',
\]
where the limits of integration $\{a,b\}$ are $\{0, x \}$ in case of Eq.(\ref{eq:less}) and $\{x, \infty \}$ in case of Eq.(\ref{eq:more}). 
Assuming that in the vicinity of $x$, $\mathcal{I} (x') \approx \mathcal{I} (x) + \mathcal{I}' (x) (x'-x) + o(x-x')$, both integrals can be estimated as
\[
\frac{|f(x')|}{|\mathcal{I}' (x)|} \exp\left[- \tau \mathcal{I} \left(x \right) \right], 
\]
i.e. in the exponential order of magnitude
\[
 \exp\left[- \tau x \mathcal{T} \left(\frac{1}{x} \right) \right] \sim \exp\left[- \tau \mathcal{I} \left(x \right) \right],
\]
from which Eq.(\ref{eq:dirlead}) follows.

\section{Asymptotics of very large deviations for Pareto cases. \label{ap:asymptotic}}

In the main text we relied on our physical intuition, and used the limiting form for $\widetilde{\psi}(s)$ and therefore of $\mathcal{L}(q)$ for large absolute values of the arguments
to calculate the Legendre transformation. Here we present some mathematical facts in support of our intuition. 

The properties of $\mathcal{L}(q) = \ln \widetilde{\psi}(-q)$ derive from those of $\widetilde{\psi}(s)$. Some of these properties are general. Thus, $\widetilde{\psi}(s) \leq 1$ and monotonically non-increasing ($\widetilde{\psi}(s)$ is completely monotonic \cite{feller}). Therefore $\mathcal{L}(q) = \ln \widetilde{\psi}(-q) < 0$ and is monotonically non-decreasing for all $-\infty < q \leq 0$. 
The asymptotics of $\mathcal{L}(q)$ follow from such of $\widetilde{\psi}(s)$. For $q \to -\infty$ the function tends to $-\infty$; for $q \to 0_-$ it tends to zero.

The Pareto distributions belong to a class of infinitely divisible distributions \cite{Thorin}. 
For these distributions $\widetilde{\psi}(s) = e^{-\phi(s)}$, where the function $\phi(s)$ is the one with completely monotonic derivative, and with $\phi(0) = 0$ \cite{feller}.
Therefore $\mathcal{L}(q) = -\phi(-q)$ is infinitely differentiable on $(-\infty, 0_-)$ and convex. 
 
For our discussions we however need only the properties of the two lower derivatives of  $\mathcal{L}(q)$, which are common for all distributions with support on the right half-line, i.e. 
for all waiting time distributions, provided these are not the delta-functions, like in simple random walks. 

We show that both $Q(q) = \mathcal{L}'(q)$ and $Q'(q) = \mathcal{L}''(q)$ are strictly positive for all $-\infty < q < 0$ (the last statement means that $\mathcal{L}(q)$ is strictly convex). 
Denoting $f(q) =\widetilde{\psi}(-q) = \int_0^\infty e^{qt}\psi(t) dt$, as in the main text, we get
$f'(q) = \int_0^\infty e^{qt}t \psi(t) dt$, $f''(q) = \int_0^\infty e^{qt}t^2 \psi(t) dt$. All three integrals are convergent for $q < 0$, the first one is always positive, and the last two 
are strictly positive for all $\psi(t) \neq \delta(t)$ (which holds in the Pareto case). 
The positivity of $Q$ follows from its representation
\[
 Q(q) = \frac{f'(q)}{f(q)},
\]
where both the numerator and the denominator are positive and finite (for $q < 0$). Turning to $\mathcal{L}''(q)$ we write
\[
 \mathcal{L}''(q) = \frac{d}{dq} \frac{f'(q)}{f(q)} = \frac{f''(q)f(q)- [f'(q)]^2}{f^2(q)}.
\]
The denominator is positive and finite. 
Now we write the expression in the numerator as a double integral:
\begin{multline*}
 f''(q)f(q)- [f'(q)]^2 = \\
 \int_0^\infty \int_0^\infty (t''^2 - t'' t') e^{q(t_1+t_2)}\psi(t'') \psi(t') dt'' dt'
\end{multline*}
and symmetrize this expression:
\begin{eqnarray*}
&& \int_0^\infty \int_0^\infty (t''^2 - t'' t') e^{q(t_1+t_2)}\psi(t'') \psi(t') dt'' dt'  = \\
&& \frac{1}{2} \bigg[\int_0^\infty \int_0^\infty (t''^2 - t'' t') e^{q(t_1+t_2)}\psi(t'') \psi(t') dt'' dt' \\
&& \qquad + \int_0^\infty \int_0^\infty (t'^2 - t'' t') e^{q(t_1+t_2)}\psi(t'') \psi(t') dt'' dt' \bigg] \\
&& = \int_0^\infty \int_0^\infty (t'' - t')^2 e^{q(t_1+t_2)}\psi(t'') \psi(t') dt'' dt'
\end{eqnarray*}
to see that it is greater than zero for all $\psi(t) \neq \delta(t-a)$. 
Therefore for our examples $\mathcal{L}''(q) > 0$ (for $-\infty < q < 0$), and thus $Q(q)=\mathcal{L}'(q)$ is a continuous, monotonically increasing
function, which thus is invertible. Moreover, $Q(q) > 0$ for all $q < 0$.

The natural variable $\xi = Q(q)= \mathcal{L}'(q)$ of the Legendre-transformed function is therefore 
a monotonic function of $q$. The inverse function $q = Q^{-1}(\xi)$ (which defines the inverse transformation $q = d \mathcal{I}/d \xi$) exists, and is monotonic in $\xi$, namely a monotonically growing function thereof. 
Since $Q'(q) = \mathcal{L}''(q)$ is strictly positive, the inverse function $Q^{-1}(\xi)$ is differentiable as well. 
The continuity and differentiability of all involved functions within their definition intervals (with exclusion of the endpoints) allows us for exchanging the limiting transitions,
i.e. to consider the limits of the Legendre transformations as Legendre-transformations of the limiting forms of the functions. 

Now we return to our Pareto distributions. For $q \to - \infty$ the value of $\xi$ tends to a positive constant $C_I$ (for type I distribution) or to zero (for type 2 distribution). 
For $q \to 0_-$ it turns either to a positive constant $\mu$ (if the mean exists), which is larger than $C_I$ or $0$ for type I and type II distributions, respectively, or diverges going to $+ \infty$ (if the mean is absent). 
The function $\mathcal{I}(\xi)$ being the Legendre transformation of $\mathcal{L}(q)$ is therefore defined either on a finite or on an infinite interval right from $\xi_{\min} \geq 0$ (being $C_I$ or zero). It is important to note that in all cases the allowed values of $\xi$ are positive: $\xi \geq 0$.

According to Eq.(\ref{eq:CTRW_gauss}) the value of $\xi$ which delivers the supremum necessary to calculate $\mathcal{X}(x)$ is given by the solution of the equation
\[
 \frac{x^2}{2}  = - \frac{d}{d\xi} \left[\frac{\mathcal{I}(\xi)}{\xi} \right]= \frac{\mathcal{I}(\xi) - \mathcal{I}'(\xi)\xi}{\xi^2},
\]
which we now consider for $x \to \infty$. The derivative of the numerator
\[
 \frac{d}{d\xi}[\mathcal{I}(\xi) - \mathcal{I}'(\xi)\xi] = - \xi \mathcal{I}''(\xi)
\]
is negative for all $\xi > 0$, so that the whole expression Eq.(\ref{eq:CTRW_gauss}) is a continuous and monotonically decreasing function of $\xi$. Therefore the solution to Eq.(\ref{eq:CTRW_gauss}),
provided it exists, is a decreasing function of $x^2$. 
In this situation, going to the limit $x^2 \to \infty$ corresponds to $\xi \to \xi_{\min}$, which on its turn corresponds to $q \to - \infty$ and $s \to \infty$.

\end{document}